\PassOptionsToPackage{table}{xcolor}
\documentclass[times]{GonnellaLab-StyleArXiv}

\usepackage{blindtext}
\usepackage[overload]{textcase}
\usepackage{amsmath,amssymb}
\usepackage{lmodern}
\usepackage{iftex}
\usepackage{enumitem}
\usepackage{pifont}
\usepackage{booktabs}
\usepackage{multicol}
\usepackage[T1]{fontenc}
\usepackage[utf8]{inputenc}
\usepackage{textcomp} % provide euro and other symbols
\usepackage{color}
\usepackage{fancyvrb}
\usepackage{natbib}
\usepackage{xcolor}
\usepackage{multirow}
\usepackage{bigdelim}
\usepackage{makecell}
\usepackage{subcaption}

\IfFileExists{upquote.sty}{\usepackage{upquote}}{}
\IfFileExists{microtype.sty}{% use microtype if available
  \usepackage[]{microtype}
  \UseMicrotypeSet[protrusion]{basicmath} % disable protrusion for tt fonts
}{}
\makeatletter
\@ifundefined{KOMAClassName}{% if non-KOMA class
  \IfFileExists{parskip.sty}{%
    \usepackage{parskip}
  }{% else
    \setlength{\parindent}{0pt}
    \setlength{\parskip}{6pt plus 2pt minus 1pt}}
}{% if KOMA class
  \KOMAoptions{parskip=half}}
\makeatother
\IfFileExists{xurl.sty}{\usepackage{xurl}}{} % add URL line breaks if available
\IfFileExists{bookmark.sty}{\usepackage{bookmark}}{}
\hypersetup{
  hidelinks,
  pdfcreator={LaTeX via pandoc}}
\DefineVerbatimEnvironment{Highlighting}{Verbatim}{commandchars=\\\{\}}
\ifLuaTeX
  \usepackage{selnolig}  % disable illegal ligatures
\fi

\begin{document}

% Please give the surname of the lead author for the running footer
\leadauthor{Lam}

\title{Collection of prokaryotic genome contents expectation rules from scientific literature}

\shorttitle{Prokaryotic genome content expectations}

% Use letters for affiliations, numbers to show equal authorship (if applicable) and to indicate the corresponding author
\author[1\space *]{Serena Lam}
\author[1,2\space \Letter,*]{Giorgio Gonnella}

\affil[1]{Institute for Microbiology and Genetics, Georg-August-Universität Göttingen, Goldschmidtstr. 1, 37077 Göttingen}
\affil[2]{Center for Bioinformatics (ZBH), Universität Hamburg, Bundesstrasse 43, 20146 Hamburg}
\affil[*]{equal contributions}

\maketitle

%TC:break Abstract
\begin{abstract}
Shaped by natural selection and other evolutionary forces, an organism's evolutionary history is reflected through its genome sequence, content of functional elements and organization. Consequently, organisms connected through phylogeny, metabolic or morphological traits, geographical proximity, or habitat features are likely to exhibit similarities in their genomes. These similarities give rise to expectations about the content of genomes within these organism groups.

Such expectations are often informally expressed in scientific literature, focusing on the analysis of individual genomes or comparisons among related groups of organisms. Our objective is to develop a system for formalized expectations as rules, facilitating automated verification, and evaluation of newly sequenced genomes.

In this study, we present a database comprising rules manually extracted from scientific literature. Furthermore, we explore the feasibility of automatizing the extraction and analysis process using large language models, such as GPT3.5 and GPT4.

We have developed a web application, \texttt{EGCWebApp}, which enables users to visualize and edit the rules. Additionally, we provided a Python library and command-line tools collection, \texttt{egctools}, to further extend the functionality for processing and managing these rules.
\end{abstract}

%TC:break main


\begin{corrauthor}
    giorgio.gonnella\at uni-goettingen.de
\end{corrauthor}

\begin{multicols}{2}
The rapid growth of available prokaryotic genomes has unraveled new possibilities for comparative genomics studies, illuminating the diversity and evolutionary relationships among these organisms \citep{CompGenProkProtocol, NGM}. These comparisons frequently reveal that organisms with shared phenotypes, habitats, or phylogenetic relationships possess common genome content—attributable to factors such as common ancestry, horizontal transfer, natural selection, and convergent evolution.

Identifying rules that express these common genetic traits based on specific factors can yield valuable insights. For instance, lineage-specific markers have been employed for quality control of genome assemblies \citep{CheckM}. Moreover, the detection of intriguing exceptions could aid in formulating hypotheses in the field of microbial genomics.

A large corpus of scientific literature describing prokaryotic genomes is available and can
serve as a source of expectation rules. In this study, we present a manually curated collection of rules, derived from scientific articles utilizing a standardized representation schema, to facilitate the automation of rule verification through computational tools.

For this purpose, we recently described a logical representation framework \citep{preprint_unambiguous} and introduced the EGC format (Expected Genome Contents), which is specifically designed for representing expectation rules concerning the content of prokaryotic genomes \citep{preprint_EGC}. We also provide a web application, EGCWebApp, allowing users to visualize and modify the rules, as well as a Python library and command-line tools collection, egctools, to further enhance the functionality for processing and managing these rules.

This work contributes to the advancement of microbial genomics research by offering a streamlined system for extracting, representing, and verifying expectation rules regarding prokaryotic genome content. Ideally, the collection could serve as a primer for a larger automated analysis of scientific literature using text mining. Thus, we explore this idea with preliminary experiments employing large language models.

\section{Methods}

\subsection{Candidate article lists}

Two different strategies were used for preparing lists of
candidate articles, of which a subset was analyzed manually.
The first strategy was used for retrieving articles related to genomes of prokaryotic organisms found in 
hydrothermal vents (data set H). Therefore, we used multiple Pubmed queries, then
listed and joined the results.
The second strategy was used for retrieving articles related to
bacterial and archaeal genomes (data sets B and A respectively). From the NCBI FTP site, we obtained the NCBI Assembly summary files for Bacteria \footnote{\url{ftp://ftp.ncbi.nlm.nih.gov/genomes/genbank/bacteria/assembly_summary.txt}} and Archaea\footnote{ \url{ftp://ftp.ncbi.nlm.nih.gov/genomes/genbank/archaea/assembly_summary.txt}}. The assembly IDs were
extracted and separated, into groups of 200 IDs each, using command line utilities.
These IDs were passed to the \texttt{esearch} utility, which retrieved the
records in the NCBI Assembly database. Then, the \texttt{elink} utility was
used for retrieving links of these records to the Pubmed database. Finally,
the \texttt{efetch} utility was used for obtaining the corresponding
Pubmed IDs.

\subsection{Selection of articles to process}

We selected a subset of the candidate article lists for manual analysis, thereby limiting our attention to those journals which were open access—either fully open access or otherwise accessible through
an existing subscription of our host institution (University of Göttingen).
We proceeded pragmatically in the subset selection
with the goal of maximizing the chances of success in extracting rules.
We focused primarily on the journals, which we considered by subjective assessment
to be more likely adherent to our goals due to the style and scope of the
included articles, however, we still surveyed articles in other journals.
In general, whenever we succeeded in finding expectations about genome contents
in articles of one journal, we progressively included more articles from it.

\subsection{Contents categorization and labeling}

To analyze the expectation extracts, we defined Groups, Genome Content Units,
Attributes and Expectations, with the definitions reported in Table \ref{tab:elements},
and further described in \citet{preprint_unambiguous}.
The information organized in this way was written in the EGC (Expectation Genome Contents) flat file format,
which we developed for this goal \citep{preprint_EGC}.

For the definition of groups of organisms and genome content units, we
selected external resources, i.e., databases and ontologies, to standardize the definitions as much as
possible (Table \ref{tab:G_types}).
Whenever a definition was not available in those resources, we adopted one of these
solutions:
\begin{itemize}
    \item Reference (DOI/PMID) to a scientific article
    \item Link to an entry in the English Wikipedia or Wiktionary
    \item Definition in plain text
\end{itemize}

\begin{table*}
\centering
\renewcommand{\arraystretch}{1.2}
\setlength{\tabcolsep}{0pt}
\begin{tabular}{p{2cm}p{0.35cm}p{3.6cm}p{0.35cm}p{1.3cm}p{4.2cm}p{0.35cm}p{6.8cm}}
\toprule
\multicolumn{3}{l}{\textbf{Group type category}} & & \multicolumn{3}{l}{\textbf{PGTO group type term}} &
\textbf{Resources for instances definition} \\
\midrule & \\[-5mm]
\rowcolor{gray!20}              & &                                          & &                           & & & \\[-3mm]
\rowcolor{gray!20}              & &                                          & & GT2001 & clade \dotfill            & & NCBI Taxonomy \citep{NCBItaxonomy} \\
\rowcolor{gray!20}              & & \multirow{-2}{=}{groups of strains \dotfill} &
                                    \multirow{-2}{=}{\ldelim\{{1}{*}}          & GT2002 & paraphyletic \dotfill     & & NCBI Taxonomy \citep{NCBItaxonomy} \\
\rowcolor{gray!20}              & &                                          & &                           &        & & NCBI Taxonomy \citep{NCBItaxonomy} \\
\rowcolor{gray!20}              & &                                          & & \multirow{-2}{=}{GT1001} & \multirow{-2}{=}{strain \dotfill}  &
                                                                                 \multirow{-2}{=}{\ldelim\{{1}{*}}     & Bacdive \citep{BacDive}            \\
\rowcolor{gray!20} \multirow{-5.7}{=}{taxon \dotfill} &
                   \multirow{-7.1}{=}{\ldelim\{{2.85}{*}} &
                      \multirow{-2.7}{=}{single strain \dotfill} &
                      \multirow{-3}{=}{\ldelim\{{1.5}{*}}                      & GT1002 & metagenome assembled \dotfill & & NCBI Taxonomy \citep{NCBItaxonomy} \\
\rowcolor{gray!20}              & &                                          & &                               & & & \\[-3mm]

\rowcolor{white}          & &                             & &                                          & & & \\[-3mm]
                          & &                             & & GH1100 & (generic) \dotfill                & & ENVO \citep{ENVO}     \\
                          & & \multirow{-2}{=}{biome \dotfill}   &
                              \multirow{-2.1}{=}{\ldelim\{{1.2}{*}}
                                                            & GH1101 & host anatomic feature \dotfill  & & UBERON \citep{Uberon} \\
                          & &                             & & GH2200 & physical-chemical \dotfill      & & MicrO \citep{MICRO}   \\
\multirow{-4}{=}{habitat \dotfill} &
  \multirow{-5.3}{=}{\ldelim\{{2.315}{*}} &
    \multirow{-2}{=}{parameter \dotfill} &
      \multirow{-2.1}{=}{\ldelim\{{1.2}{*}}                 & GH2100 & specific compound level \dotfill            & & CHEBI \citep{ChEBI}   \\
                          & &                             & &                                          & & & \\[-3mm]

\rowcolor{gray!20}          & &                            & &                                    & & & \\[-3mm]
\rowcolor{gray!20}          & &  \dotfill                  & & GP1000 & morphology\dotfill                           & & MicrO \citep{MICRO} \\
\rowcolor{gray!20}          & &                            & & GP2001 & primary metabolism \dotfill  & & MicrO \citep{MICRO} \\
\rowcolor{gray!20}          & & \multirow{-2}{=}{metabolism \dotfill} &
                                \multirow{-2}{=}{\ldelim\{{1}{*}}     & GP2002 & metabolic trait \dotfill  & & GO \citep{GO} \\
\rowcolor{gray!20}          & &                            & & GP3001 & class \dotfill                      & & OBO Relations Ontology \citep{RO} \\
\rowcolor{gray!20}          & &                            & & GP3002 & partner \dotfill          & & NCBI Taxonomy \citep{NCBItaxonomy} \\
\rowcolor{gray!20} & &
  \multirow{-3.2}{=}{biological interaction \dotfill} & \multirow{-4}{=}{\ldelim\{{2}{*}} 
                                                             & GP3003 & resultant disease/symptom \dotfill & & SNOMED CT \citep{SNOMEDCT} \\
\rowcolor{gray!20}         & &  \dotfill              & & GP4000 & taxis \dotfill                           & & MicrO \citep{MICRO} \\
\rowcolor{gray!20} \multirow{-8.8}{=}{phenotype \dotfill} & \multirow{-14}{=}{\ldelim\{{6.3}{*}}
                               &                           & &                                    & & &\\[-3mm]

\rowcolor{white}           & &                             & &                                          & & & \\[-3mm]
                           & &                               & & GL1000 & (generic) \dotfill                 & & GeoNames \citep{GEONAMES} \\
\multirow{-2}{=}{location \dotfill} & &
  \multirow{-2}{=}{geographical region \dotfill} & \multirow{-2}{=}{\ldelim\{{1}{*}}
                                                               & GL1001 & hydrothermal vent field \dotfill & & InterRidge \citep{InterRidge} \\

& \\[-4mm] 
\bottomrule
\end{tabular}
\caption{Types of organism group definition (G) 
and main external resources used for the definitions.}
\label{tab:G_types}
\vspace{5mm}
\end{table*}

\subsection{Definition of organism groups}

For a group of organisms mentioned in a text, we characterized by using a common trait,
which defines the group, and described it in EGC format \texttt{G} records.
Group type labels and their logical relationships were described
in the Prokaryotics Group Types Ontology \citep{PGTO}.
In case the group-defining common trait consists of different aspects, its definition
was divided into separate records for each of the aspects,
which were then combined. For example, Chemolithoautotrophic Archaea is a combined
group, thus it would be split into the groups: ``Chemolithoautotrophic organisms'' and ``Archaea''.
Other examples of different kinds of combinations are given in Table \ref{tab:G_derived}.

\begin{table*}
\centering
\rowcolors{2}{gray!20}{white}
\begin{tabular}{llll}
\toprule
\textbf{Type} &
\textbf{Identifier} & \textbf{Name} & \textbf{Definition} \\
\midrule
& \\[-3mm]
inverted taxon &
Gn\_Enterobacter & not Enterobacter & !Gt\_Enterobacter \\
additional spec. &
Gi\_obl\_pred & obligate predatory & derived:Gi\_predatory:obligate \\
combined taxa &
Gc\_Deltapr\_not\_Myx & non-Myxococcales δ-proteob. & Gt\_Deltaproteo \& !Gt\_Myxococcales \\
taxon/metab. &
Gc\_AOA	& ammonia-ox. Archaea	& Gm\_NH3\_ox \& Gt\_Archaea \\
taxon/disease &
Gc\_tumor\_Enterob & tumor-inducing \textit{Enterobacter}	& Gt\_Enterobacter \& Gd\_tumor \\
combined metab. &
Gc\_chm\_a\_troph & chemoautotrophic & Gp\_chm\_lth\_a\_troph $\vert$ Gp\_chm\_org\_a\_troph \\
\bottomrule
\end{tabular}
\caption{Examples of derived group definitions.}
\label{tab:G_derived}
\vspace{5mm}
\end{table*}

\subsubsection{Taxonomy}

For taxonomic groups, we employed the NCBI taxonomy database \citep{NCBItaxonomy},
which is directly linked to other resources. 
For single strains, we used either the NCBI taxonomy or the BacDive database \citep{BacDive}
whenever the strain was included in them.

\subsubsection{Habitat}

For definitions of groups organisms by their habitat, we distinguished
several cases. We defined biomes as groups of organisms living in
specific type of habitats and used the
Environment Ontology \citep{ENVO} for biome definitions.
A particular kind of biomes are organisms groups in anatomic
features of host organisms (e.g. rumen): these are not included in the ENVO, thus we used the Uberon \citep{Uberon} integrated
cross-species anatomy ontology instead.
In other cases, the habitat
is defined by a preference or requirement concerning a physical-chemical property (temperature, pH, O2 level, salinity, nutrients level). For such definitions, we employed the MicrO ontology for prokaryotic
phenotypic and metabolic characters \citep{MICRO}. 
MICRO does not cover all cases: for the preference or requirement of a single chemical compounds, we linked the definition to the term for
that compound in the Chemical Entities of Biological Interest (ChEBI) ontology \citep{ChEBI}.

\subsubsection{Phenotype}

The MicrO ontology was also employed for phenotype groups,
—such as alignment or movement towards or against a stimulus (e.g. magnetotaxis),
the reaction to Gram stain and the primary metabolims groups,
and the primary metabolism type (photo- vs.\ chemo-; litho- vs.\ organo-; auto- vs.\ hetero-).
When referring to other metabolic traits, such as the habit to oxidize
or reduce a given compound, we resorted to the Gene Ontology
\citep{GO}.

\subsubsection{Location}

For geographical regions, we employed
the GeoNames database
of names of geographical features \citep{GEONAMES}.
The database, however, did not include hydrothermal vents with appropriate
detail. Thus for this case, we instead employed the InterRidge Vents Database
\citep{InterRidge}.

\subsubsection{Biological interactions}

Groups of organisms can be defined by the type of interaction
they have with other organisms (e.g., parasites,
pathogens, symbionts) and their relative location (e.g., free-living,
episymbiont, endosymbiont, intracellular).
Many of these relations are contained in the RO, OBO Relations Ontology \citep{RO}.
Other groups were defined by describing the interaction partner
(defined by NCBI taxonomy ID), and subsequently linked together in one group along with the type of interaction.
Finally, groups can be defined by the disease or symptom resulting from the interaction,
described by terms according to the SNOMED CT,
Systematized Nomenclature of Medicine Clinical terms, Ontology \citep{SNOMEDCT}.

\subsection{Definition of genome contents}

The units of genome content (annotation features, sequence components)
mentioned in the text were characterized by their type and described
in EGC format \texttt{U} records.
The unit type labels used and their logical relationships were described
in the Prokaryotics Genome Contents Definition Ontology \citep{PGCDO}.
For orthologous group of genes, we linked to the COG and arCOG database.
For genome feature types, we used the Sequence Ontology whenever possible.
For protein families, we employed several databases, such as Interpro, 
Pfam and CDD. For enzymatic functions, we linked the EC data,
Brenda EC, and for transporters, the TC and GO.
For sets of units we employed both explicit definitions,
i.e., lists of unit names (introduced by the prefix ``set!:'') as well as
implicit definitions, where the components of the set are not listed.
The attribute modes and types or regions used in the attribute definitions
were described in the Prokaryotic Genome Contents Definition Ontology \citep{PGCDO}.

\begin{table*}
\centering
\begin{tabular}{lllp{12cm}}
\toprule
\textbf{Code} &
\textbf{Element} & & \textbf{Definition}\\
\midrule
& \\[-3mm]
\rowcolor{gray!20}
\texttt{G} & \textbf{Groups} & & the group or groups or organisms, for which the expectation is expressed \\
& \\[-3mm]
\texttt{U} & \multicolumn{2}{l}{\textbf{Genome Content Units}} & the sequence regions or annotation features, which are the subject of the expectation \\
& \\[-3mm]
\rowcolor{gray!20}
& & &\\[-3mm]
\rowcolor{gray!20}
\texttt{A} & \textbf{Attributes} &  &i.e.,\ content units and how they are considered, in particular: \\
\rowcolor{gray!20}
& & \textit{Region} & if the expectation concerns the entire genome, or portions of it\\
\rowcolor{gray!20}
& & &  (e.g.,\ only coding regions; plasmids; a specific DNA molecule; \ldots) \\
\rowcolor{gray!20}
& & \textit{Mode} & how the features are considered \\
\rowcolor{gray!20}
& & & (e.g.\ presence, count, completeness, length) \\
\rowcolor{gray!20}
& & & \\[-3mm]
\rowcolor{white}
& & & \\[-3mm]
\texttt{V/C} & \textbf{Expectations}  & & i.e.,\ a group of organisms, a genome attribute and the following elements:\\
& & \textit{Quantifier} & if the expectation applies for all organisms in the group or a portion of it\\
& & &                    (few, some, most, a given percentage, \ldots)\\
& & \textit{Operator} & the mathematical operator of the comparison (e.g., equal, larger than)\\
& & \textit{Reference} & the comparison term, i.e.,\\
& & & \ \ for \textit{value expectations} (\texttt{V}): the expected value \\
& & & \ \ for \textit{comparative expectations} (\texttt{C}): the group of organisms \\
& \\[-3mm]
\bottomrule
\end{tabular}
\caption{Logical elements of the expectation information about genome content.
The code is the record type in EGC format for the element.}
\label{tab:elements}
\vspace{5mm}
\end{table*}

\subsection{EGCwebapp}

The web application EGCwebapp — for the visualization and editing of the
contents of EGC files — was created using the Flask framework v.2.1.2 \citep{flask}.
Templates were thereby created using the Jinja2 package v.3.0.3 \citep{jinja2} and record-editing
forms were created using the wtforms package v.3.0.1 \citep{wtforms}.
For dynamic HTML content, we used the Javascript framework jQuery v.3.6.0 \citep{jquery}
and the Bootstrap library v.4.5.2 \citep{bootstrap}.
The dynamical tables were created using the Databables package v.1.10.24 \citep{datatables},
and the tooltips using Tippy v.6 \citep{tippy} and icons
from FontAwesome v.6.4.0 \citep{fontawesome}.

\subsection{egctools}

For the validation of EGC files and the collection of statistics about their contents, we created the Python package egctools. The package also handles
access to the EGC data (indexing, creation, editing) within the background of the EGCwebapp. It is based on the following open-source libraries, which we implemented and made available on Github: \texttt{egcspec}, for parsing the EGC format, using the TextFormats library \citep{textformats};
\texttt{fardes}, for parsing a micro-format for describing feature arrangements;
\texttt{lexpr}, for parsing the logical expressions used for defining combined
groups; \texttt{ec\_finder}, for suggesting EC numbers from enzyme names;
\texttt{pgcdo}, an ontology for the definition of genome contents; \texttt{pgto}, an ontology for the
definition of group of organisms.

\subsection{Text mining}

For the extraction and analysis of rules using text mining, we employed
OpenAI ChatGPT, using the gpt-3.5-turbo \citep{InstructGPT}
and gpt-4 \citep{GPT4} models. 
Since access to the API for GPT-4 was not available to the
authors, we used for both gpt-3.5-turbo and gpt-4 the conversational AI interface,
available by subscribing to Chat GPT-Plus \citep{chatgptplus}, which allowed for
unlimited interactions with the GPT 3.5 model and a maximum of 25 interactions every
3 hours with the GPT 4 model.

\subsubsection{Prompt engineering}

The prompts used for the extraction and analysis tasks are given in
Appendix 1 and are structured in the following sections:
\begin{itemize}
\item \textit{Role and Objective} naming and describing a specialized identity
for the AI conversational model and describing its general goals
\item \textit{Response guidelines} describing the specific task and
the general guidelines to follow in the answer
\item \textit{Output format} describing a structured, JSON based, format
for the outputs; variables parts of the output are given here as placeholders
\item \textit{Output values} describing the values to be included in the
variable parts of the output (instead of the placeholders)
\item \textit{Important rules} stressing the importance of adhering to the rules
and giving conventions and negative examples to avoid as numbered lists
\item \textit{Interaction} describing the interaction between user and
conversational AI, and asking for confirmation about understanding
and adhering to the defined conventions and rules
\end{itemize}

The output format was, compared to EGC, thereby simplified and
designed to not include any reference of one line to another
(Figure \ref{fig:gptoutformat}).

\begin{figure*}
\centering
(a) Text snippet extraction ChatGPT output format
\begin{verbatim}
[
  {
    "group": __GROUP__,
    "content": __CONTENT__,
    "expectation": __EXPECTATION__,
    "snippet": __SNIPPET
  }
]

\end{verbatim}

(b) Text snippet analysis ChatGPT output format
\begin{verbatim}
[
  {
    "rule":
    {
      "group":       {"name": __GROUPNAME__,
                      "type": __GROUPTYPE__,
                      "quantifier": __GROUPQUANTIFIER__}
      "content":     {"name": __UNITNAME__,
                      "type": __UNITTYPE__},
      "measurement": {"mode": __MODE__,
                      "region": __GENOMICREGION__},
      "expectaction": {"operator": __COMPARISONOPERATOR__,
                       "reference": {
                          "type": __REFERENCETYPE__,
                          "data": __REFERENCEDATA__} }
    },
    "metadata": { "explain": __REASON__, }
  }
]
\end{verbatim}
\caption{Intermediate JSON-based formats for the text mining output in (a) the sentence extraction task (b) and the sentence analysis task. The output is an array, of which here a single entry is given. Variables parts of the output are displayed here by placeholders, whose names
are written in upper case and start and end with double underscores (\texttt{\_\_}).}
\label{fig:gptoutformat}
\end{figure*}

\subsubsection{Results evaluation}

To evaluate the results of the analysis of text snippets to the manually
curated rules set, we randomly selected 30 elements of our collections of
text snippets (10 for each of the H, B and A datasets).

For the evaluation, we developed an annotation system. In particular, each
error was classified as \textit{minor} (e.g., a reasonable assumption in the output,
e.g., when not considering contexts not included in the text, but still not
completely correct) or \textit{major} (e.g., out-of-scope or invalid answers).
Then each rule in the ChatGPT output was assigned to one of the following
categories, and assigned a score accordingly:
\begin{itemize}
\item excellent: no errors (100 points)
\item good: one or few minor errors (90 points)
\item fair: multiple minor errors (60 points)
\item poor: at least one major error (20 points)
\item junk: multiple major errors (-10 points)
\item additional wrong: additional rule, not included in the manual results, since incorrect
     (-10 points).
\end{itemize}

Missing rules (present in the manual analysis but not in the output) were counted,
but no penalty was given in this case. In some cases, rules did not correspond 1:1,
e.g., a rule in the output summarized multiple rules of the manual analysis. In this
case, the rule was counted multiple times. In the opposite case, when multiple rules
in the output were summarized in the manual analysis, they were counted all together
in the score as a single rule.

\newpage

\section{Results}

\subsection{Lists of potentially genome-related scientific articles}

The first goal of our analysis is to identify scientific articles,
which could contain information about certain expectations on the content of
prokaryotic genomes.

We created 3 distinct data sets, related to bacterial genomes (B),
to archaeal genomes (A), and to genomes of prokaryotic organisms associated to hydrothermal
vents (H).

The A and B data sets were hereby created to collect articles for which the NCBI Entrez cross-database links \citep{Entrez}
included links from NCBI Assembly \citep{NCBIAssembly} to Pubmed.
The preparation of these lists on  January 26, 2022, returned 3561 papers
(B) and 190 papers (A), respectively.

The H data set was generated instead by joining four Pubmed queries, listed in Table \ref{tab:HQueries}, and contained 937 articles.

Since extraction work is very tedious, it was not feasible to work on each of the %937+190+3561
4688 papers contained in the three lists. Thus, we selected and analyzed a shorter set of the articles, using the strategy described in the Methods.
In total, we analyzed 283 articles (H: 75, A: 53, B: 155). Table \ref{tab:Journals} shows the
statistics for the rate of success in the extraction of expectation rules alongside each journal. Each of the journals were selected from a list of leading journals in which the articles were published. 

\begin{table*}
\centering
\rowcolors{2}{gray!20}{white}
\begin{tabular}{ll}
\toprule
\textbf{Query} & \textbf{N.\ results}\\
\midrule
& \\[-3mm]
\texttt{hydrothermal vent bacteria genome} & 526 \\
\texttt{hydrothermal vent bacteria sequence} & 781 \\
\texttt{hydrothermal vent archaea genome} & 170 \\
\texttt{hydrothermal vent archaea sequence} & 253 \\
\bottomrule
\end{tabular}
\caption{Queries used for the H data set collection. The number of
results displayed here includes records created on PubMed to the latest on June 15, 2022.
The overlapping sets of results from the four queries were combined into a list
of 937 unique articles.}
\label{tab:HQueries}
\vspace{5mm}
\end{table*}

\subsection{Manual collection and analysis of expectations}

From each article, we collected extracts
which expressed, directly or indirectly, expectations about the contents of genomes of prokaryotes.
We first extracted relevant tables and snippets of text, from which we thereby extracted the shortest portion
of each text, from which it was still possible to understand all relevant information.
We dissected the content of each text extract and characterized the logical components of the expressed expectations
(Table \ref{tab:elements}).

The selection criteria we used contributed to a more comprehensive set of rules. We selected sentences, which contained notable key information about a specified group of organisms. Moreover, we prioritized our selection whenever possible to statements, which assert identifiable genome features unique to a specified group of organisms, i.e., rules exclusive to that group of organisms.  We avoided vague
statements and other statements that only pertain either to a very small group of organisms or a particular isolate.
Our focus was primarily centered on gathering concise assertions with little possibility for error.

Each article was analyzed for only one of the three data sets (H, B or A).
We found expectations in 112 articles of the 238 articles analyzed (39.6\%; Table \ref{tab:data_sets}).
The next sections give an overview of the definitions of organisms groups and genome contents,
which were necessary for the rule definitions.

\begin{table*}
\centering
\rowcolors{2}{gray!20}{white}
\begin{tabular}{lllllllllll}
\toprule
\textbf{data set} &
  \textbf{Total} &
  \textbf{Processed} &
  \textbf{\%} &
  \textbf{Articles} &
  \textbf{\%} &
  \textbf{Extracts} &
  \textbf{Avg.} &
  \textbf{Rules} & 
  \textbf{Avg.} &
  \textbf{Avg.} \\
  &
  \textbf{articles} &
  \textbf{articles} &
  \textbf{total} &
  \textbf{w. expect.} &
  \textbf{processed} &
 &
  \textbf{extr./art.} &
   & 
  \textbf{rules/extr.} &
  \textbf{rules/art.} \\
\midrule \rowcolor{white}
& \\[-4mm]
H & 937 & 75 & 8.0\% & 24 & 32.0\% & 88 & 3.7 & 354 & 4.0 & 14.7 \\
B & 3561 & 155 & 4.3\% & 57 & 36.8\% & 165 & 2.9 & 464 & 2.8 & 8.1 \\
A & 190 & 53 & 27.8\% & 30 & 56.6\% & 103 & 3.4 & 309 & 3.0 & 10.3 \\
\bottomrule
\end{tabular}
\caption{Total number of scientific articles, number of processed
articles, number of articles from which expectations were found, 
number of extracted article parts (text snippets, tables),
and number of rules obtained from these extracts.
The data is shown for all three data sets.}
\label{tab:data_sets}
\vspace{5mm}
\end{table*}

\begin{table*}
\centering
\rowcolors{2}{gray!20}{white}
\begin{tabular}{l|rrr|rrr|rrr|r}
\toprule
\textbf{Journal} &
\textbf{Bacteria} & & &
\textbf{Archaea} & & &
\textbf{Hydro} & & &
\textbf{Success} \\
&
\textbf{all} &
\textbf{proc.} &
\textbf{extr.} &
\textbf{all} &
\textbf{proc.} &
\textbf{extr.} &
\textbf{all} &
\textbf{proc.} &
\textbf{extr.} &
\textbf{rate}\\
\midrule \rowcolor{white} & \\[-4mm]
Genome announc & 691 & 21 & 1 & 15 & 4 & 1 & 7 & 1 & 0 & 7.7\% \\
J Bacteriol & 521 & 17 & 8 & 38 & 13 & 9 & 24 & 2 & 1 & 56.2\% \\
Stand Genomic Sci & 174 & 7 & 3 & 21 & 7 & 3 & 13 & 1 & 0 & 40.0\%\\
Int J Syst Evol Microbiol & 169 & 10 & 0  & 16 & 5 & 2 & 142 & 5 & 0 & 10.0\%\\
PloS One & 169 & 21 & 13 & 9 & 5 & 4 & 34 & 5 & 3 & 64.5\%\\
BMC Genomics & 148 & 11 & 5 & 2 & 0 & 0 & 9 & 2 & 1 & 46.2\%\\
Proc Natl Acad Sci USA & 133 & 4 & 3 & 14 & 7 & 4 & 30 & 2 & 1 & 61.5\%\\
Appl Environ Microbiol & 81 & 2 & 0 & 8 & 1 & 1 & 89 & 12 & 3 & 26.7\%\\ 
Microbiol Resour Announc & 71 & 1 & 0 & 6 & 0 & 0 & 2 & 0 & 0 & 0.0\%\\
Front Microbiol & 61 & 2 & 1 & 3 & 0 & 0 & 43 & 2 & 0 & 25.0\%\\
Sci Rep & 45 & 0 & 0 & 1 & 1 & 1 & 13 & 0 & 0 & 100.0\%\\
Environ Microbiol & 43 & 1 & 0 & 3 & 0 & 0 & 56 & 7 & 3 & 37.5\%\\
Nature & 41 & 3 & 2 & 3 & 2 & 1 & 7 & 1 & 0 & 50.0\%\\
Science & 33 & 2 & 1 & 2 & 1 & 0 & 4 & 0 & 0 & 33.3\%\\
Genome Res & 29 & 2 & 1 & 6 & 0 & 0 & 0 & 0 & 0 & 50.0\%\\
ISME J & 24 & 3 & 3 & 6 & 2 & 1 & 50 & 4 & 2 & 66.6\%\\
mBio & 22 & 3 & 0 & 0 & 0 & 0 & 8 & 2 & 0 & 0.0\%\\
Extremophiles & 11 & 2 & 0 & 3 & 0 & 0 & 35 & 4 & 1 & 16.7\%\\
PLoS Biol & 9 & 2 & 2 & 0 & 0 & 0 & 2 & 2 & 1 & 75.0\%\\
FEMS Microbiol Ecol & 7 & 1 & 1 & 0 & 0 & 0 & 32 & 6 & 3 & 57.1\%\\
\midrule
\textit{other} &
  \textit{1079} & \textit{40} & \textit{13} &
  \textit{34} & \textit{5} & \textit{3} &
  \textit{337} & \textit{17} & \textit{5} & \textit{33.9\%} \\
\bottomrule
\end{tabular}
\caption{
For the most common journals in the data sets,
the number of total articles in the data set (all), processed articles (proc.) and
articles from which expectations were extracted (extr.).
The table includes journals which are among the leading 10 journals
with the most articles in the complete data set or the 10 journals
with the highest number of processed articles.}
\label{tab:Journals}
\vspace{5mm}
\end{table*}
%
% This table can be cited in the DISCUSSION
% to say that indeed there are different rates of success depending on the journal
%

\subsection{Definition of organism groups}

For describing the genome content expectations in the three data sets, we
defined a total of 476 groups of organisms, which in some cases were common for
more than one data set (thus we had in total 516 \texttt{G} records).
The definitions based on taxonomy alone (57.3\%), taxonomy combined with another criterion (24.2\%),
phenotype (12,0\%), habitat (5.7\%) and geographical location (0.8\%).

Most taxonomic groups were ranked as species or a higher rank (228 groups) - including the paraphyletic group Rhizobia
\citep{Rhizobia_paraphyletic}. 13 groups were ranked as strains,
or in 2 cases strain-level taxa known only because their genomes were assembled from
metagenome sequencings. For other 30 groups, the definition criterion was the non-membership to a taxon 
or set operations (unions and/or intersection) of different taxa.
Most phenotype groups were defined by aspects of their metabolism
(29 groups; Table \ref{tab:G_by_metabolism}), such as source of energy, reducing equivalents
and organic carbon, or the ability or inability to use some chemicals.
In 26 phenotype groups (Table \ref{tab:G_by_interaction}), the definition criteria were based on aspects of biological interaction
with other organisms (i.e., class of interaction, partner, resultant disease or symptom) and their combinations.
Finally, the other two definitions were one for Gram-staining and one for the reaction to a stimulus (i.e., magnetotaxis).
Habitat groups were defined by the name of the biome, e.g. bathypelagic waters
(9 groups; Table \ref{tab:G_habitat_kind}), by the preference or requirement of given physical-chemical environmental parameters
(16 groups; Table \ref{tab:G_by_requirement}), or the ability
to thrive under laboratory conditions (2 groups).
Finally, 3 groups were defined as organisms living in a given geographical region or specific geographical feature.

\begin{table*}
\centering
\rowcolors{2}{gray!20}{white}
\begin{tabular}{lll}
\toprule
\textbf{Identifier} & \textbf{Name} & \textbf{Definition} \\
\midrule
& \\[-3mm]
\texttt{Gpm\_pht\_troph } & phototrophic & MICRO:0001457 \\
\texttt{Gpm\_lth\_troph } & lithotrophic & MICRO:0001459 \\
\texttt{Gpm\_chm\_troph } & chemotrophic & MICRO:0001458 \\
\texttt{Gpm\_chm\_lth\_troph } & chemolithotrophic & MICRO:0001476 \\
\texttt{Gpm\_chm\_lth\_a\_troph } & chemolithoautotrophic & MICRO:0001477 \\
\texttt{Gpm\_chm\_org\_a\_troph } & chemoorganoautotrophic & MICRO:0001480 \\
\texttt{Gpm\_a\_troph } & autotrophic & MICRO:0001456 \\
\texttt{Gpm\_h\_troph } & heterotrophic & MICRO:0001474 \\
\texttt{Gpm\_obl\_a\_troph } & obligate autotrophic & derived:Gpm\_a\_troph:obligate \\[2mm]
\midrule
\texttt{Gm\_photosyn } & photosynthetic & GO:0015979 \\
\texttt{Gm\_purple\_photosyn } & purple photosynthetic & Wikipedia:Purple bacteria \\
\texttt{Gm\_asaccharolytic } & asaccharolytic & Wiktionary:asaccharolytic \\[2mm]
\midrule
\texttt{Gm\_NH3\_ox } & ammonia-oxidizing & GO:0019329 \\
\texttt{Gm\_NO2\_ox } & nitrite-oxidizing & GO:0019332 \\
\texttt{Gm\_S\_ox } & sulfur-oxidizing & GO:0019417 \\
\texttt{Gm\_H2S\_ox } & sulfide-oxidizing & GO:0019418 \\
\texttt{Gm\_SO4\_red } & sulfate-reducing & GO:0019419 \\
\texttt{Gm\_Fe2\_ox } & Fe(II)-oxidizing & GO:0019411 \\
\texttt{Gm\_CH4\_ox\_aer } & aerobic methane-oxidizing & doi:10.1007/978-1-4020-9212-1    \_139 \\
\texttt{Gm\_ac\_ox } & acetic acid oxidizing & MICRO:0000773 \\[2mm]
\midrule
\texttt{Gm\_CH4\_gen } & methanogenic & MICRO:0000166 \\
\texttt{Gm\_ac\_gen } & acetogenic & Wikipedia:Acetogenesis \\
\texttt{Gm\_H\_troph } & hydrogenotrophic & Wikipedia:Hydrogenotroph \\
\texttt{Gm\_1C\_troph } & methylotrophic & Wikipedia:Methylotroph \\
\texttt{Gm\_CH4\_troph } & methanotrophic & Wikipedia:Methanotroph \\
\texttt{Gm\_S\_troph } & thiotrophic & doi:10.1007/978-1-4020-9212-1  \_206 \\
\bottomrule
\end{tabular}
\caption{List of organism group definition (G) records
         by primary metabolism or single metabolic traits.
         Note: doi:10.1007/978-1-4020-9212-1  is the Encyclopedia
         of Geobiology \citep{EnGeobiology}.}
\label{tab:G_by_metabolism}
\vspace{5mm}
\end{table*}

\begin{table*}
\centering
\rowcolors{2}{gray!20}{white}
\begin{tabular}{lll}
\toprule
\textbf{Identifier} & \textbf{Name} & \textbf{Definition} \\
\midrule
& \\[-3mm]
\texttt{Gh\_marine } & marine waters & ENVO:01000320 \\
\texttt{Gh\_mesopelagic } & mesopelagic & ENVO:00000213 \\
\texttt{Gh\_bathypelagic } & bathypelagic & ENVO:00000211 \\
\texttt{Gh\_marsed } & marine sediment & ENVO:00002113 \\
\texttt{Gh\_hydrvents } & marine hydrothermal vents & ENVO:01000122 \\
\texttt{Gh\_soil } & soil & ENVO:01001044 \\
\texttt{Gh\_subsurface } & subsurface & ENVO:01001776 \\
\texttt{Gh\_gut } & gut/enteric & ENVO:2100002 \\
\texttt{Gh\_rumen } & rumen & UBERON:0007365 \\
& \\[-3mm]
\bottomrule
\end{tabular}
\caption{List of organism group records (G) defining a biome.}
\label{tab:G_habitat_kind}
\vspace{5mm}
\end{table*}

\begin{table*}
\centering
\rowcolors{2}{gray!20}{white}
\begin{tabular}{llll}
\toprule
\textbf{Requirement kind} &
\textbf{Identifier} & \textbf{Name} & \textbf{Definition} \\
\midrule
& \\[-3mm]
temperature
& \texttt{Gr\_mesophilic } & mesophilic & MICRO:0000111 \\
& \texttt{Gr\_thermophilic } & thermophilic & MICRO:0000118 \\
& \texttt{Gr\_hyperthermophilic } & hyperthermophilic & MICRO:0001294 \\
& \texttt{Gr\_psychrophilic } & psychrophilic & MICRO:0001306 \\
& \texttt{Gr\_psychrotrophic } & psychrotrophic & OMP:0005006\\
& \\[-3mm]
\midrule
oxygen level
& \texttt{Gr\_aerobic } & aerobic & MICRO:0000494 \\
& \texttt{Gr\_anaerobic } & anaerobic & MICRO:0000495 \\
& \texttt{Gr\_microaerophilic } & microaerophilic & MICRO:0000515 \\
& \\[-3mm]
\midrule
salinity
& \texttt{Gr\_halophilic } & halophilic & MICRO:0001314 \\
& \\[-3mm]
\midrule
pH
& \texttt{Gr\_neutrophilic } & neutrophilic & MICRO:0001546 \\
& \texttt{Gr\_acidophilic } & acidophilic & MICRO:0001390 \\[1mm]
\midrule
nutrients level
& \texttt{Gr\_copiotrophic } & copiotrophic & Wikipedia:Copiotroph \\
& \texttt{Gr\_oligotrophic } & oligotrophic & Wikipedia:Oligotroph \\[1mm]
\midrule
specific nutrient
& \texttt{Gcl\_urearich } & urea-rich & CHEBI:16199 \\
& \\[-3mm]
\bottomrule
\end{tabular}
\caption{List of organism group records (G) defining organisms with specific preferences or requirements for physico-chemical parameters and specific chemical compounds.}
\label{tab:G_by_requirement}
\vspace{5mm}
\end{table*}

\begin{table*}
\centering
\rowcolors{2}{gray!20}{white}
\begin{tabular}{lll}
\toprule
\textbf{Identifier} & \textbf{Name} & \textbf{Definition} \\
\midrule
& \\[-3mm]
\texttt{Gic\_hashost } & host-associated & RO:0002454 \\
\texttt{Gic\_intracellular } & intracellular & RO:0002640 \\
\texttt{Gic\_pathogenic } & pathogenic & RO:0002556 \\
\texttt{Gic\_predatory } & predatory & RO:0002439 \\
\texttt{Gic\_obl\_intracell } & obligate intracellular & derived:Gic\_intracellular:obligate \\
\texttt{Gic\_obl\_pred } & obligate predatory & derived:Gic\_predatory:obligate \\
\texttt{Gic\_fac\_pred } & facultative predatory & derived:Gic\_predatory:facultative  \\
\texttt{Gic\_free } & free-living & OMP:0007646 \\
\texttt{Gic\_episymbiont } & episymbiont &  Wiktionary:episymbiont\\
\texttt{Gic\_endosymbiont } & endosymbiont & Wikipedia:Endosymbiont \\[2mm]
\midrule
\texttt{Gip\_siphonaptera } & interacting with Siphonaptera (fleas) & taxid:7509 \\
\texttt{Gip\_vertebrata } & interacting with Vertebrata (vertebrates) & taxid:7742 \\
\texttt{Gip\_Calyptogena } & interacting with Calyptogena & taxid:6589 \\
\texttt{Gip\_Apompejana } & interacting with Alvinella pompejana & taxid:6376 \\[2mm]
\midrule
\texttt{Gio\_Lyme } & Lyme disease & sctid:23502006 \\
\texttt{Gio\_relapsfever } & relapsing fever & sctid:420079008 \\
\texttt{Gio\_tumor } & tumor & sctid:108369006 \\
\bottomrule
\end{tabular}
\caption{List of organism group definition (G) records
         by biological interaction class, partner and outcome for the partner.}
\label{tab:G_by_interaction}
\vspace{5mm}
\end{table*}

\subsection{Definition of genome contents}

Since our goal is to describe expectations of genome contents, we must define those contents first.
In our system, a measurable property of the genome is called \textit{genome attribute}. It refers
to a \textit{genome content unit} (GCU), which is a single sequence unit or annotation feature
or a set of those. The GCU is observed in the whole genome, or part of it, and in a given modality (e.g., presence/absence, count, sequence length), which can be absolute or relative to another
GCU (e.g., relative frequency of a given COG to all COG annotations).

Thus, GCUs are first defined (as records of type \texttt{U} in the EGC format), then
used for the definition of attributes (as records of type \texttt{A}).
For specific genes or proteins, additional information can be added to support the correct
identification, such as links to protein family or domain models (as records of type \texttt{M}).
This was done for 192 GCUs, to which we added 637 M records (668 when counting duplicates among the data sets). Table \ref{tab:modeldb} shows a list of the databases employed in those records.

\begin{table*}
\centering
\rowcolors{2}{gray!20}{white}
\begin{tabular}{lll}
\toprule
\textbf{Database} & \textbf{Models} & \textbf{Reference} \\
\midrule
      InterPro & 431 & \citet{InterPro} \\
      Pfam & 139 & \citet{Pfam} \\
      PROSITE & 51 & \citet{Prosite} \\
      TIGRFAMs & 55 & \citet{TIGRFAMs} \\
      PIRSF & 14 & \citet{PIRSF} \\
      SMART & 11 & \citet{SMART} \\
      SFLD & 1 & \citet{SFLD} \\
      SUPfam & 1 & \citet{SUPFAM} \\
      PRINTS & 1 & \citet{PRINTS} \\
      HAMAP & 1 & \citet{HAMAP} \\
\bottomrule
\end{tabular}
\caption{Number of models for each database used in M records}
\label{tab:modeldb}
\end{table*}

\subsubsection{Genome content units}

We defined a total of 657 genome content units. As for groups definitions, some were common
to different data sets, therefore we defined a total of 691 \texttt{U} records.
The units belong to 3 distinct categories: ``single units'',
i.e., specific units identified by a name or symbol,
which were considered as ``indivisible'' (42.6\% of the definitions);
``categories'' (40.0\%), to which multiple single units belong (corresponding to a logical OR operation); and ``sets'' (17,4\%), which consisted of multiple single units which altogether
define the unit (corresponding to a logical AND operation).

Single units include specific genes (234 units) and proteins (39 units).
They do not include heteromeric protein complexes,
for which distinct sub-units can be considered singularly, and are thus defined as sets.
Other single unit definitions are used for implementing our system sequence statistics, such as the genome length—they define DNA bases and amino acids (7 units).
Categories at protein level includes homologs of specific proteins (7 units),
families, clans and protein domains (49 units), which are identified by a
Pfam \citep{Pfam}, InterPro \citep{InterPro}, CDD \citep{cdd} or TC \citep{tcdb} accession, and functional categories (90 cases) which are mostly defined
by a EC number or GO term \citep{GO}. Categories at gene level include homologs of specific genes (8 units), cluster
of ortholog genes and categories of clusters (82 units), mostly defined by COG \citep{COG} and arCOG \citep{arcog} accessions.
Finally, features types, which were defined whenever possible by a SO term, account for 22 unit definitions.
Our definition system also allows for defining categories by specifying their members (e.g., multiple specific genes),
which was used in 5 other unit definitions.
In definitions of sets, the components of the set can be enumerated and described as other unit records (45 sets), or instead be given only a set name or a description (69 sets).
Sets definitions at protein level include protein complexes (21 definitions) and metabolic pathways, which
can be seen as sets of enzymatic functions (27 definitions). Sets definitions at gene level 
include operons (10 definitions), gene clusters (25 definitions), gene systems,
i.e., genes with a common function, although not always spatially clustered (18 definitions)
and generic sets of multiple genes defined by enumerating their members (1 definition).
Other sets include heterogeneous types of features, such as genomic islands (2 definitions)
and feature arrangements (10 definitions).

\subsubsection{Genome attributes}

The attribute definitions refer to a GCU and include the genomic region and
mode of measurement.
We defined 530 distinct attribute records (in total 547). Although the same GCU is used sometimes in different attribute definitions (23 GCUs), the number is lower than that of GCUs, since some GCUs are not the focus of an attribute definition (152 GCUs) and are instead used by other GCU definitions (e.g., when enumerating the members of a set) or as reference GCUs in attributes with relative measurement modes.

Table \ref{tab:U_with_A_modes} reports the number of attributes and the measurement modes for each category of GCU.
For single units and categories, the following modes were used: presence/absence (310 attributes), count (106 attributes), total or average sequence length (2 attributes). Thereby, count and sequence length can be either absolute (50 attributes) or relative to a second specified GCU (53 attributes).
For set units, it must also be specified if the measurement concerns the complete set or single members. In the former case, the modes were ``presence of a complete set'' (106 attributes), and ``count of number of complete instances of the set'' (3 attributes). In the latter case, in which a set becomes
equivalent to a category with the same members, the mode is ``presence of any member of the set'' (8 attributes).

\begin{table*}
\centering
\begin{tabular}{llllll}
\toprule
\textbf{Category} & \textbf{Type} & \textbf{N.GCUs} &
\textbf{Resources} & \textbf{N.attributes} & \textbf{Modes} \\
\midrule
& \\[-3mm]
\multirow{ 4 }{*}{\textit{ simple }(280)}& specific\_gene & 234      &&  111 &\scriptsize{\makecell{presence: 102\\count: 9\\} }\\\cmidrule{2-6}& specific\_protein & 39      &&  35 &\scriptsize{\makecell{presence: 29\\count: 6\\} }\\\cmidrule{2-6}& amino\_acid & 4      &
                  \scriptsize{\makecell{SO: 4\\} }
                &  3 &\scriptsize{\makecell{count\_relative: 3\\} }\\\cmidrule{2-6}& base & 3      &&  1 &\scriptsize{\makecell{count: 1\\} }\\

\midrule
& \\[-3mm]
\multirow{ 11 }{*}{\textit{ category }(263)}& function & 90      &
                  \scriptsize{\makecell{EC: 43\\GO: 5\\BRENDA\_EC: 1\\\textit{none}: 41\\} }
                &  87 &\scriptsize{\makecell{presence: 79\\count: 7\\count\_relative: 1\\} }\\\cmidrule{2-6}& ortholog\_group & 74      &
                  \scriptsize{\makecell{COG: 40\\arCOG: 30\\\textit{none}: 4\\} }
                &  77 &\scriptsize{\makecell{presence: 40\\count\_relative: 37\\} }\\\cmidrule{2-6}& family\_or\_domain & 49      &
                  \scriptsize{\makecell{Pfam\_clan: 22\\Pfam: 12\\InterPro: 12\\CDD: 2\\TC: 1\\} }
                &  48 &\scriptsize{\makecell{presence: 31\\count: 17\\} }\\\cmidrule{2-6}& feature\_type & 22      &
                  \scriptsize{\makecell{SO: 12\\\textit{none}: 10\\} }
                &  21 &\scriptsize{\makecell{presence: 11\\count: 8\\length\_relative: 1\\average\_length: 1\\} }\\\cmidrule{2-6}& ortholog\_groups\_category & 8      &
                  \scriptsize{\makecell{COG\_category: 8\\} }
                &  8 &\scriptsize{\makecell{count\_relative: 8\\} }\\\cmidrule{2-6}& gene\_homologs & 8      &&  8 &\scriptsize{\makecell{presence: 8\\} }\\\cmidrule{2-6}& protein\_homologs & 7      &&  7 &\scriptsize{\makecell{presence: 7\\} }\\\cmidrule{2-6}& +specific\_gene & 2      &&  2 &\scriptsize{\makecell{presence: 2\\} }\\\cmidrule{2-6}& +ortholog\_group & 1      &
                  \scriptsize{\makecell{COG: 1\\} }
                &  0 &\\\cmidrule{2-6}& +base & 1      &&  3 &\scriptsize{\makecell{count\_relative: 3\\} }\\\cmidrule{2-6}& +unit & 1      &&  1 &\scriptsize{\makecell{count: 1\\} }\\

\midrule
& \\[-3mm]
\multirow{ 10 }{*}{\textit{ set }(114)}& gene\_cluster & 25      &&  23 &\scriptsize{\makecell{complete\_presence: 23\\} }\\\cmidrule{2-6}& metabolic\_pathway & 22      &
                  \scriptsize{\makecell{KEGG: 5\\\textit{none}: 17\\} }
                &  26 &\scriptsize{\makecell{complete\_presence: 25\\members\_presence: 1\\} }\\\cmidrule{2-6}& protein\_complex & 19      &&  18 &\scriptsize{\makecell{complete\_presence: 13\\members\_presence: 3\\presence: 1\\complete\_count: 1\\} }\\\cmidrule{2-6}& gene\_system & 18      &&  18 &\scriptsize{\makecell{complete\_presence: 17\\complete\_count: 1\\} }\\\cmidrule{2-6}& arrangement & 10      &&  10 &\scriptsize{\makecell{complete\_presence: 10\\} }\\\cmidrule{2-6}& operon & 10      &&  10 &\scriptsize{\makecell{complete\_presence: 9\\complete\_count: 1\\} }\\\cmidrule{2-6}& +metabolic\_pathway & 5      &
                  \scriptsize{\makecell{KEGG: 2\\\textit{none}: 3\\} }
                &  5 &\scriptsize{\makecell{complete\_presence: 4\\members\_presence: 1\\} }\\\cmidrule{2-6}& genomic\_island & 2      &&  2 &\scriptsize{\makecell{complete\_presence: 2\\} }\\\cmidrule{2-6}& +protein\_complex & 2      &&  4 &\scriptsize{\makecell{complete\_presence: 2\\members\_presence: 2\\} }\\\cmidrule{2-6}& +specific\_gene & 1      &&  2 &\scriptsize{\makecell{complete\_presence: 1\\members\_presence: 1\\} }\\

\bottomrule
\end{tabular}
\caption{Number of genome content unit definitions by type (simple, category,
set) and unit, with the number of attributes defined for the units, and next to that, different measurement modes.}
\label{tab:U_with_A_modes}
\vspace{5mm}
\end{table*}

\subsection{Rules of expectation}

In total, our collection includes 1152 expectation rules about contents of prokaryotic genomes.
All rules were based one at least one source from our data set
of document extracts (text snippets and tables) and in 74 cases, more then one.

From these, most rules (1071) are expressed in absolute terms,
i.e., as comparisons to a reference value.
In most cases, the attributes employed boolean measurement modes (presence of a GCU),
and thus the rules consisted in a equality to True or False value (944 rules).
In 37 cases, the value rule compares an attribute value to a single numerical
reference value with numerical binary comparison operators, such as 
equal or less than. In 5 other cases, numerical range limits for the expected
attribute value are given. In 83 cases, a qualitative level (e.g. "high level")
without exact specification is given, and in 2 cases an approximate value
(e.g., "approximately 0.02") is given.

The other 81 rules are expressed as comparisons of two groups of organisms.
In these rules, the operator was, in 39 cases, a numerical comparison operator
(larger than or smaller than), while, in the other 42 cases,
an additional specification detail was given (``much'' larger than, ``slightly'' larger than).

Most rules apply to all members of a group, but our system additionally allows  registration of
rules which apply only to a portion of the members; in this
case, a group quantifier is employed, i.e., "most" (used in 72 rules), "some" (used in 54 rules),
"many" (33 rules), "rarely" (4 rules) or a rough percentage of the group portion (12 rules).    

\subsection{Tools for handling the rules collections}

Our expectation rule collections were stored in a file format created for
this goal, called EGC (Expected Genome Content) and described in
the manuscript \citet{preprint_EGC}. In order to facilitate
working with these files, we created the open source Python package egctools
(available at \url{https://github.com/ggonnella/egctools}), provided command line tools and access to the format from
Python, and a web application, egcwebapp (available at \url{https://github.com/ggonnella/egcwebapp}), for comfortable visualization
and editing of the information.

\subsubsection{The egctools package}

The egctools package includes command line
tools for common operations, such as extracting all lines in the file
related to a selected one (\texttt{egctools-extract}),
creating content reports (\texttt{egctools-stats} and \texttt{egctools-table}) or merging multiple files while skipping duplicate lines
(\texttt{egctools-merge}). An example output of \texttt{egctools-extract} is given
 in Figure \ref{fig:egctoolsmerge}.
The Python package parses the format
using the
Textformats \citep{textformats} specification for EGC, contained
in the \texttt{egc-spec} Github repository, as well as the fardes
arrangements description mini-format \citep{fardes},
the lexpr package for parsing logical expressions in derived
group definitions \citep{lexpr},
and the ontologies PGTO, which organizes the hierarchy of organism group types
\citep{PGTO} and PGCDO, which supports the definition of genome contents \citep{PGCDO}.

\begin{figure*}
\begin{footnotesize}
\begin{verbatim}
[272]	G	Gt_TMS	TMS	taxonomic	taxid:933	TR:Z:genus
[282]	    G	Gc_HV_TMR_TMS	members of HV, TMR and TMS	combined	Gt_HV & Gt_TMR & Gt_TMS
[280]	    G	Gc_CLA_TMS	chemolithoautotroph TMS	combined	Gpm_CLA_troph & Gt_TMS
[273]	    G	Gc_H2S_ox_CLA_TMS	sulfide-ox. CLA TMS	combined	Gt_TMS & Gm_H2S_ox & Gpm_CLA_troph
[994]	    V	VH309	SH101	Ac_p_RubisCO_IAc	Gt_TMS:most	==	True
[558]	        A	Ac_p_RubisCO_IAc	Up_RubisCO_IAc	count
[328]	             U	Up_RubisCO_IAc	specific_protein	.	.	carboxysomal RubisCO (form IAc)
[110]	        S	SH101	PMID:29521452	        Genomes of most members of TMS encode only ...
[20]	             D	PMID:29521452	doi/10.1111/1462-2920.14090        
[995]	    V	VH310	SH101	Ap_g_cbbM	Gt_TMS:most	==	False
[607]	        A	Ap_g_cbbM	Ug_cbbM	presence
[400]	            U	Ug_cbbM	specific_gene	.	cbbM	form II RuBisCo large subunit (cbbM)
[502]	                M	Ug_cbbM	InterPro	IPR020871	RuBisCO_lsuII	UP:Z:Q59462
\end{verbatim}
\end{footnotesize}
\caption{Example output of egctools-extract, visualizing the records interconnected
with the \texttt{Gt\_Thiomicrospira} group definition in the bacterial data set.
The numbers on the left correspond to line numbers in the input file. The lines relationships
are made clear by different indentation levels.
For ease of representation, some of the contents are shortened here: e.g.,\ \emph{Thiomicrospira}
is abbreviated to TMS, \emph{Thiomicrorhabdus} to TMR, \emph{Hydrogenovibrio} to HV, and chemolithoautotroph to CLA.}
\label{fig:egctoolsmerge}
\end{figure*}

\begin{figure*}
    \centering
    \begin{subfigure}[t]{0.55\textwidth}
        \centering
        \includegraphics[width=\linewidth]{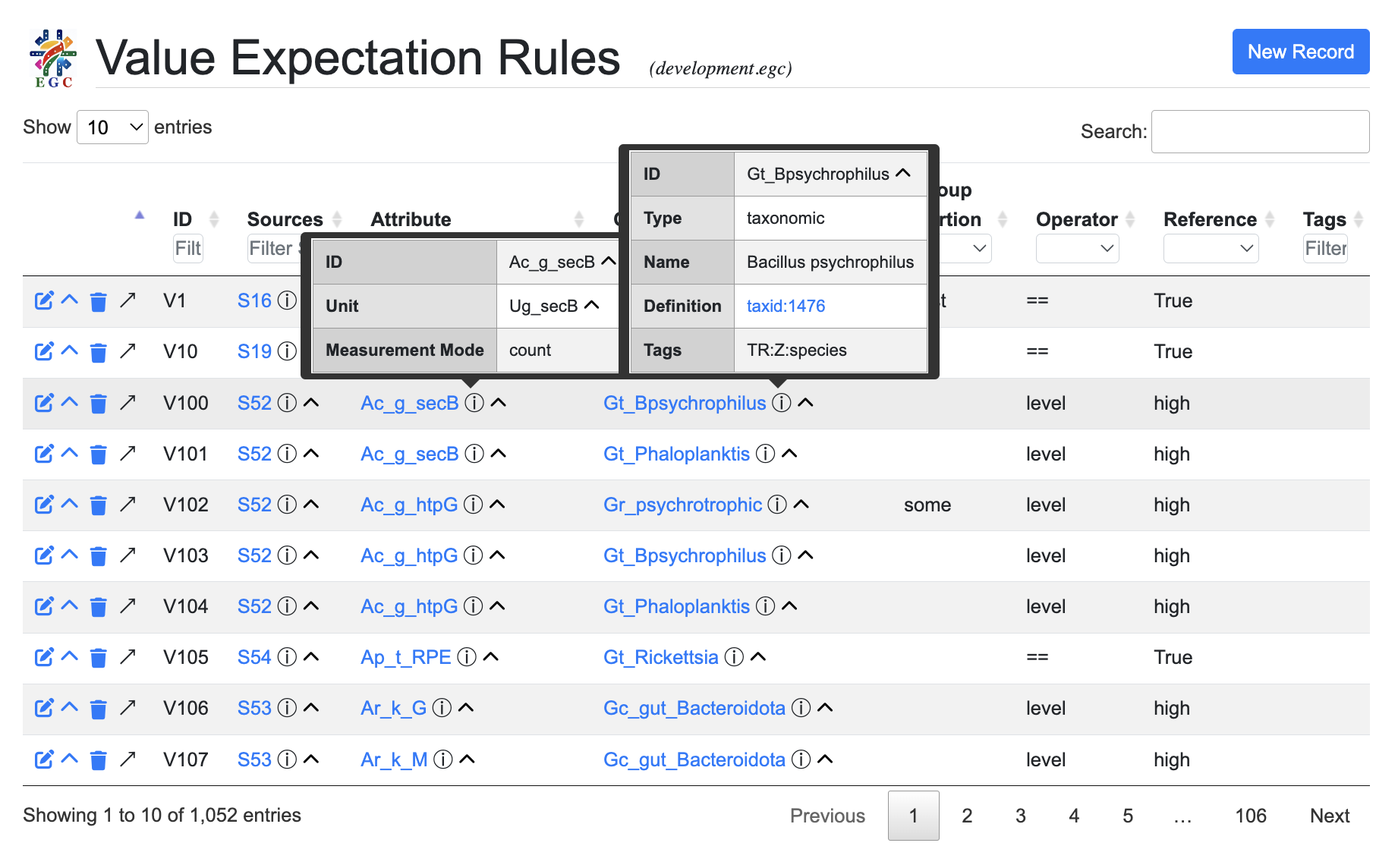} 
        \caption{Value expectations table} \label{subfig:vrules}
        \vspace{1cm}
    \end{subfigure}
    \hfill
    \begin{subfigure}[t]{0.4\textwidth}
        \centering
        \includegraphics[width=\linewidth]{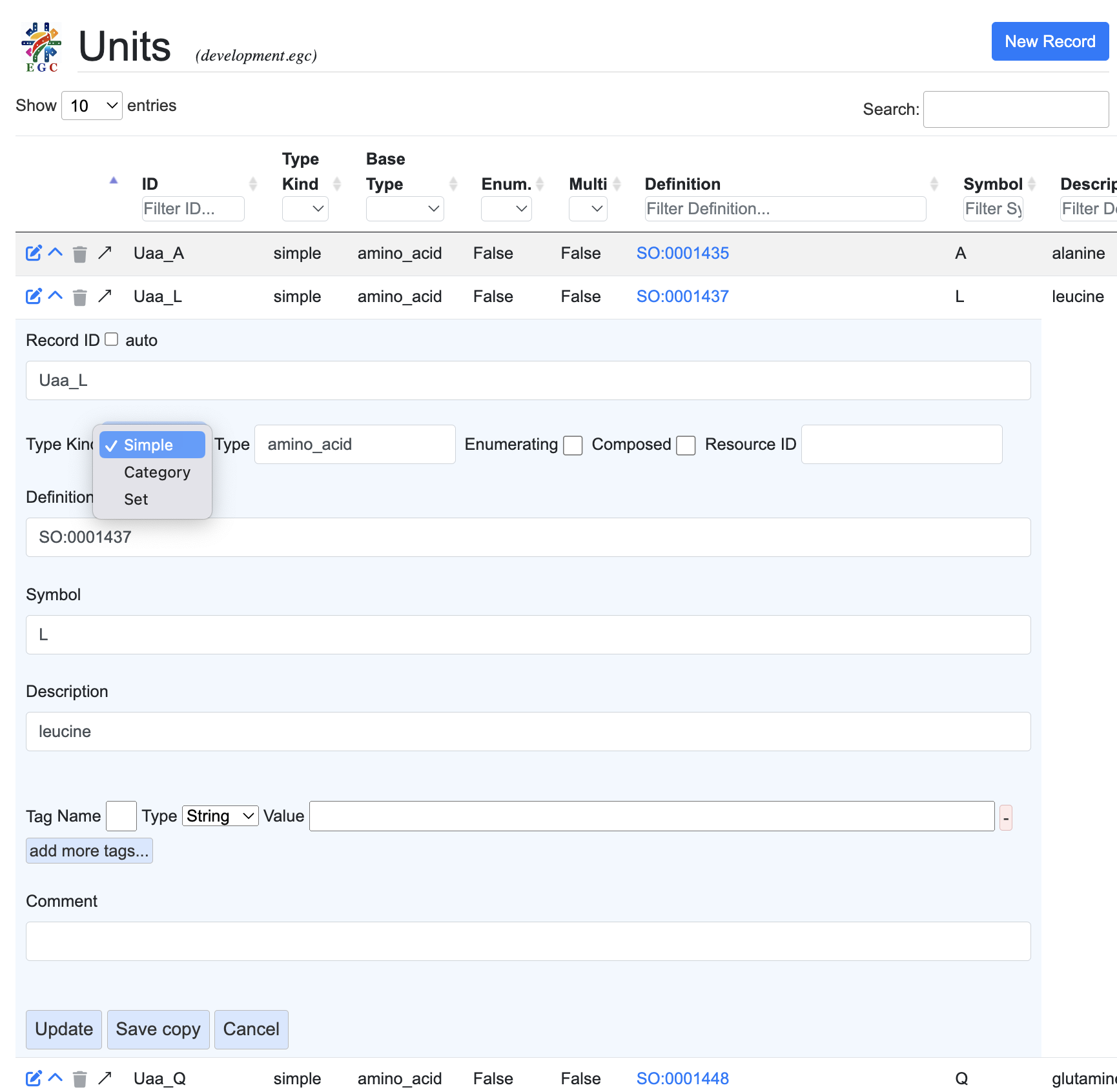}
        \caption{Editing a GCU record} \label{subfig:U-editing}
    \end{subfigure}
    \vspace{2cm}
    \begin{subfigure}[t]{0.55\textwidth}
        \centering
        \includegraphics[width=\linewidth]{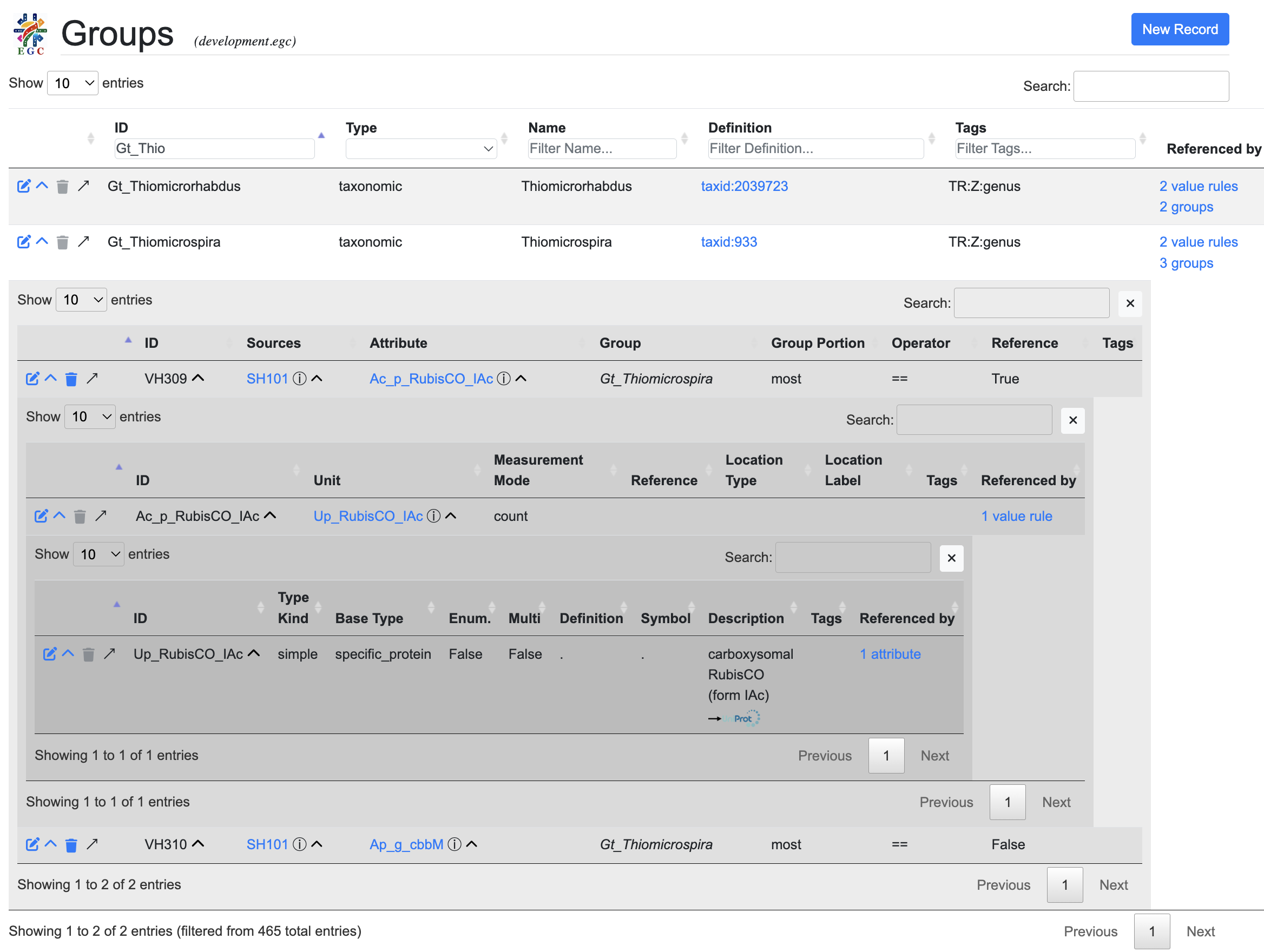}
        \caption{Table nesting} \label{subfig:nested}
    \end{subfigure}
    \hfill
    \begin{subfigure}[t]{0.4\textwidth}
        \centering
        \includegraphics[width=\linewidth]{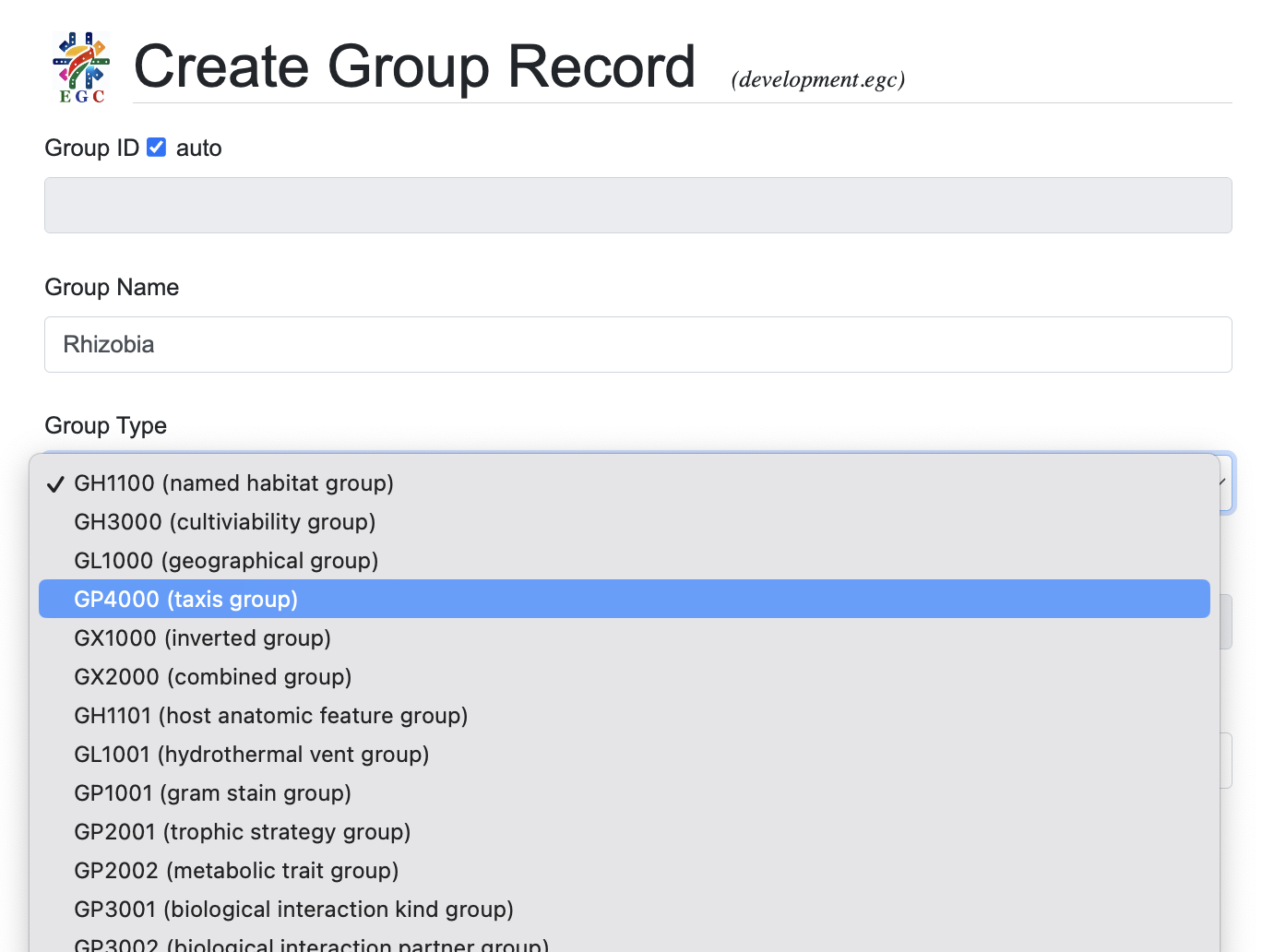} 
        \caption{Creating a group} \label{subfig:pgto}
    \end{subfigure}
 %   \begin{subfigure}[t]{0.3\textwidth}
 %       \centering
 %       \includegraphics[width=\linewidth]{egcwebapp-A-editing.png}
 %       \caption{Editing an attribute} \label{subfig:A-editing}
 %   \end{subfigure}
  \caption{Overview of the web application EGCWebapp.
  The entry points for the data exploration are pages with tables of
  records by type, e.g. value expectation rules. In these, contents
  can be sorted and filtered, and tooltips such as in Fig.\ \ref{subfig:vrules}
  show information about the connected records. Records can be edited in forms which appear
  dynamically inline in the table views (Fig. \ref{subfig:U-editing}).
  Clicking on references opens nested tables, which allows traversing the connection graph,
  (Fig.\ \ref{subfig:nested}). 
  Validations and lists of values,
  such as the group types defined in PGTO (Fig. \ref{subfig:pgto}) make
  it easier to create new record definitions. Record creation,
  editing and deletion and file saving are only available if read-write mode is enabled.}
  \label{fig:egcwebapp}
\end{figure*}

\begin{figure*}
\end{figure*}

\subsubsection{The egcwebapp web application}

The web application, egcwebapp, is based on egctools for reading
and storing the data,
as well as handling cross-references between records and other functions, 
such as automatic suggestions of record identifiers.
Figure \ref{fig:egcwebapp} shows screenshots of some of the application
functions.

When opening the application with the browser, a file can be uploaded
to the server and its content can be visualized.
The application provides a main menu, which allows dynamic switching back-and-forth between
tables of documents, document extracts, organism groups, genome content units, models, attributes, value rules and comparative rules.
The tables are also dynamic and allow for pagination, sorting and filtering/searching by different criteria.

Whenever a record refers to an external database or ontology, a link
is provided to the corresponding entry in the web site
of that resource. Many records are also interconnected to each
other (e.g., a rule refers to an attribute, which refers to one or multiple
GCUs). The contents of the referred records are visualized as
tooltips when hovering over the record IDs in the tables. Furthermore,
both refer to records, and records which refer the current one can
be opened as nested tables. The system allows dynamic traversal of the interconnected graph without leaving the current page,
but the focus reset at any time to any of the visualized records of the initial page.

The web application can be used both in read-only mode for visualizing data
of an existing data set, or in read-write mode, which can be used as a graphical editor for EGC
files. New records can be created from the table pages. Optionally, unique,
compact and informative IDs are automatically assigned to the records.
Validations are used to verify that the edited records are correctly defined.
Values for categorical variables (e.g., group types) are selected
from the available ontology terms, with the possibility to add new
values if none of the existing is adequate.
Records which are not interconnected to other records can be deleted.
Existing records can be edited in nested forms. 
Finally, the edited file
can be saved back on the local system.

\subsection{Extending the collection by text mining: preliminary evaluation}

To investigate a possible extension of the rules collection using automated text mining,
we designed prompts for the extraction of sentences, using our logical framework, from a portion of the
text of a scientific article, as well for the analysis of extracted snippets
of text. The prompts used for these tasks are given in Appendix 1
and further details are given in the methods.

The first task that we tried to automatize is the extraction of informative snippets of text,
explicitly stating or implying one or multiple expectations about the contents of prokaryotic
genomes. To each interaction, we provided to the chat the largest possible part of the text,
which was accepted by the chat interface, usually comprising one or multiple sections of the paper.
We did not include tables, figure captions, the methods sections and the references.

At the current stage of development, the prompt for this task retrieves a higher number of text snippets than
the manual analysis. A reason for this is that rules regarding single strains and species were generally
not included in the latter. Since ChatGPT cannot distinguish single strains or species,
those rules are incorrectly included in its output.
Furthermore, the answers include wrong snippets of text, which do not contain expectations
about the genome contents of prokaryotic organism groups.

The second task, for which we tested the use of ChatGPT, was the analysis of short snippets of text,
extracted from scientific papers, using our logical framework. This task was separated from the
generic extraction task, because of the limited output size and to simplify the results evaluation.
In fact, we could compare the results of the analysis to our manually curated data set, by
providing a subset of the manually extracted snippets of text. Using the evaluation system
described in the Methods section, we classified and assigned a score to each rule in the output.
The results of the evaluation using GPT3.5 and GPT4 is given in Table \ref{tab:GptEval}.

\begin{table*}
\centering
\rowcolors{2}{gray!20}{white}
\begin{tabular}{l|rrrrc|rrrr}
\toprule
           &  \multicolumn{4}{c}{\textbf{GPT 3.5}}
                             & & \multicolumn{4}{c}{\textbf{GPT 4}} \\
\midrule
           &  H  &  B  &  A  & \textit{sum} & \ \ &  H  &  B  &  A & \textit{sum} \\
\midrule
N.extracts & 10  & 10  & 10  & 30  & & 10  & 10  & 10 & 30 \\
N.rules    & 50  & 23  & 31  & 104 & & 50  & 23  & 31 & 104 \\
\midrule
excellent  &  2  &  5  &  1  &   8 & & 15  &  8  &  8 &  31 \\
good       &  1  &  5  &  8  &  14 & & 20  &  7  &  3 &  30 \\
fair       &  5  &  2  &  3  &  10 & &  0  &  2  &  8 &  10 \\
poor       & 21  &  6  &  7  &  34 & &  0  &  0  &  3 &   3 \\
junk       &  2  &  0  &  0  &   2 & &  2  &  0  &  0 &   2 \\
\midrule
missing    & 19  &  5  & 12  &  36 & & 13  &  6  &  9 &  28 \\
add.wrong  &  1  &  9  &  3  &  13 & &  2  &  3  &  2 &   7 \\
\midrule
score      & 980  & 1100 & 1110 & 3190 & & 
            3260  & 1520 & 1590 & 6370 \\
average    & 19.6 & 47.8 & 35.8 & 30.7 & & 
             65.2 & 66.1 & 51.3 & 61.3 \\
\bottomrule
\end{tabular}
\caption{
  Number of rules of expectations in each evalutation
  category (excellent, good, fair, poor, junk) and number
  of missing rules, as well as additional wrong rules
  output by GPT3.5 and GPT4 for each of the datasets
  (H, B, A). The score row contains a score, computed
  as follows: for each excellent 100 points, good 90,
  fair 60, poor 20, junk -10, missing 0, and additional wrong
  -10 points. The average score (last row) is computed
  dividing the score by the number of rules and it is equal
  to the percent of the maximum achievable score (i.e.
  all excellent answers, without additional wrong answers).
}
\label{tab:GptEval}
\end{table*}

\newpage

\section{Discussion}

In the process of identifying and extracting text snippets from scientific
literature and analyzing the
information contained therein, we encountered various challenges and made several operational
decisions to ensure the quality and relevance of our results. In this section, we discuss and
defend these decisions, highlighting their implications for our analysis method and results.

\subsection{Selection of the documents for the manual analysis}

Due to the significant effort required to perform manual analysis of scientific articles, we were
unable to consider all scientific articles included in the initial lists.
To increase the efficiency of the selection
process, we focused our attention on papers from journals, from which we had already selected some
articles, and were successful in finding expectations about the contents of genomes.
By doing so, we were able to gradually identify more articles that were likely to contain relevant information, while minimizing the time and resources
required to complete the manual analysis. Although this approach may have limited the scope
of our analysis to some extent, it was a necessary compromise due to the limited number of project resources.

\subsection{Inclusion of non-taxonomic organism groups}

At first glance, it may seem appealing to only consider groups of organisms which
are related phylogenetically, since we are analysing genome contents. Genome contents are
predominantly shaped by the genomes of the ancestors of those organisms and
evolutionary forces (e.g., natural selection and genetic drift).
For example, some gene ortholog groups are present only in specific lineages \citep{LSPFV}.

However, other factors, other than phylogeny, can account for the expected content of genomes.
One of these factors are the characteristics of the habitat.
Even if ``everything'' might indeed be ``everywhere'' \citep[see e.g.][]{Gonnella2016}, it is debatable,
how much allopatric speciation is relevant to bacteria \citep{allopatric},
as stated by Bass-Becking, the environment selects \citep{BBBreal}.
Thus, analyses of microbiomes would likely reveal that the most
prominent members of an environment will share common characteristics.
Furthermore, microorganisms living in the same location often exchange genetic material
by horizontal transfer, shaping their genome evolution \citep{HGTweb, HGT_gut}.

For those reasons, the criteria for defining groups of organisms were not only
based on phylogeny, but also on phenotype, habitat and geographic location.

\subsection{General exclusion of information on single strains or species}

As a general rule, we preferred excluding data about single strains or species in our analysis.
Although such information is often available, it is also the least helpful for
our goal, which is to provide general information about the genomes of groups of organisms,
which could be falsified for some members of the group.

Still, some data for single strains and species are included, either because those
groups were used for the definition of larger groups, or because some traits were
asserted to be exclusive for those group, thus actually giving information about the
absence of those traits in the much larger (and therefore interesting) set of all organisms,
which are \textit{not} members of any group.

\subsection{Reliance on definitions from external resources}

Our representative system for the expectation rules (EGC format)
was designed in mind to never lose sight of the goal, i.e., the verification,
when including new data. Therefore, most of the definitions
given in the rule data sets are connected to external resources, such as biological
databases and ontologies.

For example,
the NCBI Taxonomy database \citep{NCBItaxonomy} was selected as the preferred source
for our taxonomic group definitions
instead of the List of Prokaryotic names with Standing in Nomenclature \citep{LPSN}.
Although, the NCBI Taxonomy database sets out a disclaimer that it shall not be considered an authoritative
source of taxonomy information, it is, in fact, the resource which is mostly
referred to by other genomics databases provided by NCBI, e.g.,\
NCBI assembly \citep{NCBIAssembly}, but also by other institutions.

As a general rule, despite there being no absolute standards for referring
to data in biology, some databases are favored, and thus selecting these
allows the user to obtain the data necessary for verification. 
At the same time, the representation system is agnostic to external
resources, thus users can use other databases or ontologies, if desired.

Using external resources is not always successful, for instance in our case, we sometimes had difficulties or failures finding an adequate definition despite checking multiple databases.
In such cases, we either linked a reference work (such as Wikipedia or Wiktionary),
or a scientific article. In other cases, we provided a concise free-text definition, to which a verification of the rule will at first require some additional work
to identify organism groups or genome content units which reflect the definitions.

\subsection{Non-fully defined information}

In some cases, the natural language, employed in the scientific articles, allows using
expressions which are not able to be unambiguously translated into a specific data point.
For example, an assertion, that the count of a certain category of genes is high
in a given group of organisms, does not specify the exact definition of ``high''.
Nevertheless, we decided to include these types of assertions into our analysis. The underlying
idea for this decision is that the user, who wants to verify a rule, can provide
the missing information at that stage. For example, the user can decide that,
within a specific context, a high level means higher than 10 copies, while in another
context, the threshold could be 2 copies.

Several features in the format enables such fuzzy definitions. In the example above,
a value rule could employ the operator \texttt{level} and the value can be \texttt{high} or \texttt{low}. This was the case in 83 rules (7.2\% of all rules) in our data sets.
In comparative rules, the operator can include an intensity level.
Accordingly \texttt{<<} and \texttt{>>} are used for indicating ``much'' lower or ``much'' higher
values, \texttt{~<} and \texttt{~>} for ``slightly'' lower or ``slightly'' higher, and
\texttt{~=} for a ``similar'' value (44 rules, or 3.8\% of all rules, in our data sets). All these operators require the application,
which wants to verify rules, to indicate the specific breadth of intensity, e.g., what is``much'' higher than what and what, instead,
is higher, but not necessarily considered ``high''.
Similarly, genome content unit definitions can be based upon other unit definitions
by providing an additional specifier, which can be fully defined
(e.g., ``larger than 100 kb''), or also generic
(e.g., ``large''). The latter case, which occurs only once in our data sets,
would require a further definition at the time of verification.

\subsection{Non-universal rules}

Another context, in which fuzzy information is sometimes employed, is the optional
group quantifier, which defines the portion of the group, for which a rule holds.
For example, in a text it could be stated that a given gene is present in ``some''
or ``most'' of the members of a group, or a rough percentage could be given
(say ``more than 90\%'').

These quantifiers were used in 174 rules, i.e., in 15,1\% of all rules in our data sets.
During verification, the user would need to consider that due to the inherent flexible nature of these types of rules, that even if the rule
does not holds true, the rule could still be an interesting fact, but to a lesser degree than for the stricter rules which hold true for all members of the group.

\subsection{Text mining}

The collection of manually curated rules extracted from scientific literature presented in this manuscript
represents a proof-of-concept, derived from only a subset of the available literature. In fact, we were
able to manually analyse only about 6\% of the papers, which we retrieved in our candidate lists.
Also, it is very likely that other publications, not included in those list, contain expectations about
prokaryotic genomes.

In order to provide a more comprehensive collection of rules, an automated procedure, based on text mining,
would be needed. Recently, OpenAI made the large language models GPT-3.5 and GPT-4 available to the public.
These models, which are among the most powerful currently available, are useful for a variety of tasks
\citep{GPT4}. Thus, we evaluated if they can be used for our goal, of extending the collection.

The preliminary analysis presented in this paper showed encouraging results. Thereby, we crafted prompts
for the extraction and analysis of snippets of text expressing expectations from scientific articles.
The evaluation of the results showed that the analysis, using our logical framework, succeeded in about 1/3
(GPT-3.5) or 2/3 (GPT-4) of all cases (Table \ref{tab:GptEval}). A simplified JSON output therefore was produced, since attempts to produce
the full EGC format often provided invalid output, e.g., with dandling references to not existing lines (Figure \ref{fig:gptoutformat}).

At the current state of this work, further manual work would be needed to filter the extracted sentences,
to evaluate the correctness of the analysis and to convert the results to EGC. However, these tasks could
be automated by further improving the system. The accuracy could be further improved by using a fine-tuning model
based on the manually curated dataset.

\section{Conclusion}

In this manuscript, we present a collection of rules of expectations about the
contents of prokaryotic genomes. Each rule was compiled by extracting it
from a part of a peer-reviewed scientific article. The purpose of this collection is to
provide a set of previously published assertions, which can be verified or falsified,
whenever new data or analysis methods become available.
The collection can be useful for different goals: e.g., verification of the quality
of new data (assuming the rules hold), or conversely, conception of new theories when new,
discordant data become available.  

To make the rules verifiable, these are not only provided as
a set of text snippets extracted from literature. These are instead analyzed logically
alongside other entities or conditions found also in the text, this includes
the organism groups, the genome contents, the way these contents are measured (which we
call attributes) and the comparison—including reference values or reference organism groups—assessment.

The collection described in this manuscript can be considered a proof-of-concept.
It is important to note that a much larger scale of analysis of the scientific literature
would be required to fully capture the available information.

Future studies could, thus, aim to expand the collection.
Since this work is very time-consuming and requires extensive domain knowledge,
they could be supported by the use of large language models (LLMs) such as GPT-4 \citep{GPT4}.
LLMs have demonstrated impressive abilities in language understanding and could be
used to identify relevant scientific articles, extract rules from them,
and even generate new rules based on existing data.
Here we demonstrate how the use of GPT models achieved encouraging results
in extracting expectations from scientific texts.
Once this system is refined,
it could greatly facilitate the expansion and update of our collection of rules over time.

\begin{acknowledgements}
Giorgio Gonnella has been supported by the DFG Grant GO 3192/1-1 ‘`Automated characterization of microbial genomes and metagenomes by collection and verification of association rules’’. The funders had no role in study design, data collection and analysis, decision to publish, or preparation of the manuscript.
\end{acknowledgements}

\begin{contributions}
Giorgio Gonnella conceived the project, acquired the funding, created the candidate lists of scientific articles, created the logical
analysis framework and representation system, implemented the Python libraries and the web application, checked and corrected the
extracted rules, performed the preliminary text mining analysis using ChatGPT and wrote the original draft of the publication.
Serena Lam performed the manual rules extractions from the scientific papers, reviewed and edited the manuscript.
\end{contributions}

\begin{interests}
 The authors declare no competing financial interests.
\end{interests}

\bibliography{references}

\end{multicols}

\newpage

\section{Appendix 1: Prompts for ChatGPT}

\subsection{Common prompt sections}

The prompts used for different tasks (extracting snippets of text
from scientific papers and analysing extracted snippets using our framework) had common sections, which are given here.

\subsubsection{Preamble}

The common preamble of the prompts, defining an identity and objective
was the following:

\begin{verbatim}
# Role and Objective

Your identity is EgcGPT, an AI conversational model specializing in the extraction
of text segments from scientific literature that convey expectations regarding
the genomic contents of prokaryotic organism groups. These expectations could be
either a comparison of a specific genomic measure with a benchmark value or a
comparison with a different organism group's corresponding value.
\end{verbatim}

\subsubsection{Epilogue}

The end of the prompts, giving instructions about the output, explaining the mechanism of interaction, recommending to adhere to the rules and conventions and asking for a confirmative answer, was also common to the different tasks, and is given here:

\begin{verbatim}
The JSON output is well-formatted, indented, and enclosed within triple backticks
(```json       ```). This will ensure the answer is properly displayed as Code.
There will be nothing else in the answer, no comments, no explanations, just the
JSON code in the code block.

# Interaction

If you understand all the requirements, respond with "EgcGPT: yes" and nothing else.
In the following interactions, I will provide a text input per round, and you will
analyze it and provide an appropriate response as EgcGPT. Please strictly adhere to
all the above-stated conventions, requirements, and output formats.
\end{verbatim}

\subsection{Extraction of text snippets}

The following prompt was used for the extraction of text snippets (single or multiple sentences) from an  input text (one or multiple sections of a scientific paper). The text followed the common preamble and was followed by the common epilogue given above.

\begin{verbatim}
# Response Guidelines

Subsequent inputs containing texts for analysis will be delivered in following
interactions. EgcGPT's responses should be formatted in JSON, embodying an array
of entries. In instances where no such genomic expectation is present in the input
text, the response should be an empty array. Typically, an input text contains
only few expectations. Avoid extracting sentences that lack any of the defined
components or do not unequivocally represent genomic expectations for prokaryotic
groups, which must be more extensive than a single species or strain.

The text segments extracted by EgcGPT are succinct yet informative. They should
constitute the shortest text portion encapsulating all necessary components. These
components include the organism group, the genome content type and name, the
anticipated content value or a comparison with another organism group.
Any expectation not related to genome content or any text mentioning genome content
without associating it with an organism group should be excluded from the response.

# Output format

The format of the each entry in the answer of EgcGPT is given below, included in
triple quotes ```. The format is given here with the required indent and order of
the keys. There are no other keys in the answer. Thereby identifiers enclosed by
double underscores __ indicate the variables, strings for which EgcGPT will choose
appropriate values, following the rules and explanations given in the next section.
\end{verbatim}

\newpage

\begin{verbatim}
```JSON
{
 "group of organisms": __GROUP__,
 "genome content": __CONTENT__,
 "expectation": __EXPECTATION__,
 "text extract": __EXTRACT__
}
```

# Output values

__GROUP__ is a group of prokaryotic organisms explicitly mentioned in the extracted
text segment (__EXTRACT__). The string cannot be empty or "not applicable"; the
group must comprise more than a single species or strain; if the __EXTRACT__ does
not contain any name of a group, the array entry is not included in the output;

__CONTENT__ is a content of the genome (sequence, annotation feature or product of
the genome) explicitly referred to in the extracted text segment; if __EXTRACT__
does not mention any genome content, the array entry is not included in the output;

__EXPECTATION__ is the expectation about the content in the group, as explicitly
contained in the extracted text;

__EXTRACT__ is an unmodified segment of the input text, comprising single or
multiple complete sentences, as short as possible, but still containing the
necessary references to __GROUP__, __CONTENT__ and __EXPECTATION__.

# Important Rules

The following rules are always strictly followed by EgcGPT:

(1) each output array entry is a table containing exactly four entries:
    "groups of organisms", "genome content", "expectation", and "text extract".
(2) All expectations must pertain to organism groups and genome contents.
(3) The extracted text snippets contain the name of the organism and reference
    to a specific genome content.
(4) Expectations regarding organism groups that are presumably single species or
    strains are not included in the output.
(5) Accuracy takes precedence over the number of sentences extracted

\end{verbatim}

\subsection{Text snippet analysis}

The following prompt was used for the text snippet analysis task from an 
input text (short text snippet from a scientific paper). The text followed the common preamble and was followed by the common epilogue given above.

\begin{verbatim}
# Response Guidelines

The input texts will be given in the next chat iterations. EgcGPT answers in JSON
format. The JSON represents an array of entries. Each entry describes an expectation 
about the contents of genomes of prokaryotic genomes which can be deduced from the
input text and gives some metadata. Sometimes an input text does not imply any valid
rule, then the answer will be an empty array.

\end{verbatim}

\newpage

\begin{verbatim}

# Output format

The format of the each entry in the answer of EgcGPT is given below, included in
triple quotes ```. The format is given here with the required indent and order of
the keys. There are no other keys in the answer. Thereby identifiers enclosed by
double underscores __ indicate the variables, strings for which EgcGPT will choose
appropriate values, following the rules and explanations given in the next section.

```JSON
{
    "rule": {
      "group":       {"name": __GROUPNAME__, "type": __GROUPTYPE__,
                      "quantifier": __GROUPQUANTIFIER__}
      "content":     {"name": __UNITNAME__,  "type": __UNITTYPE__},
      "measurement": {"mode": __MODE__,      "region": __GENOMICREGION__},
      "expectaction": {"operator": __COMPARISONOPERATOR__,
                       "reference": {"type": __REFERENCETYPE__,
                                     "data": __REFERENCEDATA__} }
    },
    "metadata": { "explain": __REASON__, }
}
```

# Output values

"group":  the group of organisms for which the expectation about the contents holds

   __GROUPNAME__: how the group is called in the input text

   __GROUPQUANTIFIER__: if any, an expression from the input text used next to the
      group name to indicate a portion of the group, e.g. many, none, some, or a
      percentage (e.g. 90\%). In some other cases it is understandable from the
      context that the rule does not hold for all elements of the group and
      __GROUPQUANTIFIER__ is set to an expression, e.g. "some",  which explains for
      which part of the group it holds. In many cases there will be no such
      expression in the input text or be implied from the context, and
      __GROUPQUANTIFIER__ is then set to a dot "."

   __GROUPTYPE__: describes the type of criterion which defines the group;
                  it must be one of the following values:
       (1) "taxon": the group is a taxon
       (2) "habitat_kind": the group are organisms which live in a given habitat;
                       which is referred to by name, e.g. bathypelagic
       (3) "habitat_requirement": group of organism which share a preference or
                              requirement regading the habitat, e.g. acidophils
       (4) "location": the group are organisms which live in a given specific
                   geographical location; it is not a location in terms of habitat
                   type (e.g. "marine bacteria", which would be "habitat_kind")
                   and is not a location in terms of interaction with other organisms
                   (e.g. "intracellular bacteria", which would be "interaction")
       (5) "morphology": the group is defined by a morphological trait
                         (e.g. reaction to Gram stain, presence of a flagellum)
       (6) "metabolism": the group is defined by a metabolism trait
       (7) "interaction": the group is defined by the kind of interaction
               to other organisms (eg. pathological), the place of interaction
               (e.g. intracellular), the organism partner of the interaction
               (e.g. Homo sapiens) and/or the consequences of the interaction for
               the partner (e.g. cancer).
       (8) "taxis": the group is defined by the reaction to a stimulus
               (e.g. magnetotaxis)

 "content":
    the unit of genome content (genome feature, set of genome features,
    product of the genome, sets of products, etc) for which the expectation is
    described

    __UNITNAME__: the name of the unit, as it is referred to in the input text
    __UNITTYPE__: describes the type of content unit
                  it must be one of the following values:
        (1) "gene": if the unit is a specific gene
        (2) "protein": if the unit is a specific protein
        (3) "family": if the unit is a family of proteins
        (4) "domain": if the unit is a protein domain
        (5) "orthologs": if the unit is a group of ortholog genes
        (6) "function": if the unit is a function, e.g. an enzymatic or transport
             activity
        (7) "pathway": if the unit is a metabolic pathway
        (8) "gene_system": if the input text refers to the unit as a "gene sytem"
        (9) "gene_cluster": if the input text refers to the unit as a "gene cluster"
        (10) "island": if the input text refers to the unit
             as a "genomic island" or "gene island"
        (11) "operon": if the input text refers to the unit as an "operon"
        (12) "feature_type": if the unit is a type of feature,
             e.g. "protein-coding gene"
        (13) "arrangement": if the unit is an order of features in the genome

  "measurement":
      the kind of measurement of the unit for which the expectation is described

    __MODE__: the mode of measurement, it must be one of the following values:
      (1) "count" if the rule is based on the number of instances of a unit
      (2) "presence" if the rule describes presence or absence of a unit
      (3) "length" if the rule is about the sequence length of a unit

    __GENOMICREGION__: the region of the genome to consider; one of:
      (1) "." if no specific region of the genome is mentioned in the input text
      (2) name of a genomic region mentioned in the input text

 "expectation":
    the expectation about the value of "unit" measured as described by "measurement"
    in the genomes of the organisms belonging to "group"; be as specific as possible,
    based on the input text

    __COMPARISONOPERATOR__: the comparison operator of the expectation;
      it must be one of the following, depending on the input text and on the
      value of __MODE__:
      __MODE__ presence: "=="
      __MODE__ length/count: one of the following:
        (1) "==" equal
        (2) "~" roughly equal
        (3) ">=" equal or larger than
        (4) ">=~" equal or slightly larger than
        (5) ">~" slightly larger than
        (6) ">" larger than
        (7) ">>" much larger than
        (8) "<=" equal or smaller than
        (9) "<=~" equal or slightly smaller than
        (10) "<~" slightly smaller than
        (11) "<" smaller than
        (12) "<<" much smaller than
        (13) "><" in the specified range of values
        (14) "<>" out of the specified range of values
        (15) "$" has the given level (high, low, etc)

    __REFERENCETYPE__: is either "group" or "value"
    __REFERENCEDATA__:
      if __REFERENCETYPE__ is "group": name of the group of organisms to compare to
      if __REFERENCETYPE__ is "value": value (or value range) to compare to

"explain":
  __REASON__ is a very concise and direct explanation of why the expectation can be
  inferred from the input text. Thereby you must avoid verbose formulations such as:
  "the text implies/states/explains/describes that FACT" and instead just write FACT.
  Keep it as short as possible. It does not include information outside the scope of
  the rule, such as organisms groups or genome contents not mentioned elsewhere in
  the output.

# Important Rules

The following rules are always strictly followed by EgcGPT:

    (1) __GROUPNAME__ is not "." and not empty
    (2) __UNITNAME__ is not "." and not empty
    (3) __GROUPTYPE__ is one of the 8 group types mentioned above
    (4) __UNITTYPE__ is one of the 13 unit types mentioned above
    (5) __MODE__ is one of the 3 modes mentioned above
    (6) __COMPARISONOPERATOR__ is one of the 14 operators mentioned above
    (7) __REFERENCETYPE__ is either "group" or "value"
    (8) if __REFERENCETYPE__ is "group", then __REFERENCEDATA__ is a group name;
        else if __REFERENCETYPE__ is "value", then __REFERENCEDATA__ is one of the
        following: a single numeric value, two numeric values indicating a range,
        a boolean (true, false) or a quantifier (a word expressing a quantitative
        level, such as high or low, no other kind of word);
    (9) if __MODE__ is "presence", then __REFERENCETYPE__ is "value", the operator
        is "==" and __REFERENCEVALUE__ is either true or false; if __MODE__ is
        "count" or "length", then __REFERENCEVALUE__ is not a boolean (true, false)
    (10) if __COMPARISONOPERATOR__ is "<>" or "><", then __REFERENCETYPE__ is
         "value" and __REFERENCEVALUE__ is a pair of numeric values expressing the
         range limits; else if __COMPARISONOPERATOR__ is not "<>" or "><",  then
         __REFERENCEVALUE__  is not a pair of numeric values
    (11) if __COMPARISONOPERATOR__ is "$", then __REFERENCETYPE__ is "value" and
         __REFERENCEVALUE__ is a word expressing quantitative level, such as high
         or low; else if __COMPARISONOPERATOR__ is not "$", then __REFERENCEVALUE__
         is not a word expressing  a quantitative level
    (12) __REASON__ is concise and does not contain formulations such as "in the
         text", "in the input text", "the text implies/explains/describes that"
         or similar

\end{verbatim}

\end{document}